\documentclass[aps,preprint,11pt]{revtex4}

\usepackage{amssymb,epsf}
\usepackage{epsfig}
\usepackage{epstopdf}
\usepackage{graphicx}
\usepackage{color}

\begin{document}

\title{\textbf{\Large Properties of the defect modes in 1D lossy photonic crystals containing two types of negative-index-material defects}}
\author{A. Aghajamali $^{1,2}$\footnote{email address:
alireza.aghajamali@fsriau.ac.ir} M. Hayati $^{3}$\footnote{email
address: mjhayati@yahoo.com}, C.-J. Wu $^{4}$\footnote{email
address: jasperwu@ntnu.edu.tw}, and M. Barati $^{2}$\footnote{email
address: barati@susc.ac.ir}}

\affiliation{\small $^1$ Young Researchers Club, Science and Research Branch, Islamic Azad University, Fars, Iran\\
$^2$ Department of Physics, Science and Research Branch, Islamic Azad University, Fars, Iran\\
$^3$ Department of Physics, Marvdasht Branch, Islamic Azad University, Marvdasht, Iran\\
$^4$ Institute of Electro-Optical Science and Technology, National Taiwan Normal University, Taipei 116, Taiwan, R.O.C.\\
}

\begin{abstract}
In this paper, the characteristic matrix method is employed to
theoretically investigate the propagation of electromagnetic waves
through one-dimensional defective lossy photonic crystals (PCs)
composed of negative index materials (NIMs) and positive index
materials (PIMs). We consider symmetric and asymmetric geometric
structures with two different types of NIM defect layers at the
center of the structure. The effects of the polarization and the
angle of incidence on the defect modes in the transmission spectra
of both structures are investigated. The results show that the
number of the defect modes within the photonic band gap (PBG)
depends on the type of the NIM defect layer and is independent of
the geometrical structure. Moreover, it is shown that the defect
mode frequency increases as the angle of incidence increases. This
property is also independent of the geometry of the structure. The
results can lead to designing new types of narrowband and
multichannel transmission filters.
\end{abstract}\

%\pacs{04.50.+h, 04.20.Jb, 04.70.Bw, 04.70.Dy}
\maketitle

\section{Introduction}
Over the past two and half decades, a new class of material called
photonic crystals (PCs) has emerged. The PCs originate from
theoretical work of Yablonovitch, and John's experimental work,
which were published almost simultaneously in 1987
\cite{re1,re2}.The PCs are artificial dielectric or metallic
structures in which the refractive index changes periodically in
space. This kind of periodic structure affects the propagation of
electromagnetic waves in the similar way as the periodic potential
in a semiconductor crystal affects the electron motion by defining
allowed and forbidden electronic energy bands. Whether or not
photons propagate through PC structures depends on their frequency.
Frequencies that are allowed to travel are known as modes, and
groups of allowed modes form bands. Disallowed bands of frequencies
are called photonic band gaps (PBGs) \cite{re3,re4}. These PBGs are
also called the Bragg gaps because they originate from the Bragg
scattering in the periodic structure. The properties of PBGs in
one-dimensional (1D) PCs have been proven to play an important role
in some potential applications such as photonic devices, optical
filters, resonance cavities, laser applications, high reflecting
omnidirectional mirrors, and the optoelectronic circuits
\cite{re5,re6,re7,re8}.

By breaking the periodicity of the conventional PC structure, we
will have a defective crystal. This can be performed by changing
physical parameters, such as changing the thickness of one of the
layer, adding another medium to the structure, or removing a layer
from PCs \cite{re9,re10,re11,re12,re13,re14}. By introducing a layer
with different optical properties, localized defect modes, which are
also called resonant transmission peaks, can be generated within the
PBG due to the change of the interference behavior of light
\cite{re14,re15}, very similar to the defect states that are
generated in the forbidden band of doped semiconductors. Upon
prediction of the existence of materials with negative refractive
index (NRI) in 1968 by Veselago \cite{re16}, such materials, which
have negative permittivity and permeability simultaneously, have
received extensive attention for their very unusual electromagnetic
properties. Negative refractive index materials, or simply negative
index materials (NIMs), are now also known as metamaterials.
Recently, with the possibility of producing metamaterials, PCs with
metamaterials, called metamaterial photonic crystals (MPCs) have
been made.

In several papers the properties of the defect modes in different 1D
conventional PC and 1D MPC structures have been investigated by
introducing positive or negative indices defects
\cite{re17,re18,re19,re20,re21,re22,re23,re24,re25,re26,re27,re28,re29,re30,re31,re32,re33,re34,re35}.
Following the interesting report by Wu et al. \cite{re23} on the
properties of the defect modes in 1D PCs with symmetric and
asymmetric geometric structures, here we investigate the properties
of the defect modes in the transmission spectra of 1D defective
symmetric and asymmetric lossy MPCs which are composed of negative
refractive index material defect layer at the center of the crystal.
The outline of this paper is as follows. In Section 2, two geometric
MPC structures, the permittivity and permeability of two types of
NIMs, and also the characteristic matrix method and its formulation
are presented. The numerical results and discussions are given in
Section 3, and the conclusion is presented in Section 4.

\section{MPC structure and characteristic matrix method}

The 1D defective MPCs, which are constituted by alternative layers
of NIM and positive index material (PIM), under study with
asymmetric and symmetric structures in air with a defect layer at
the center of the structures are shown in Figures 1(a) and 1(b),
respectively, where NIMs are dispersive and dissipative. We assume
that layers \emph{A} and \emph{C} (defect layer) are NIMs, and layer
\emph{B} is a PIM. \emph{N} is the number of the lattice period, and
also $d_{i}$, $\varepsilon_{i}$, and $\mu_{i}$ ($i=A,B,C$) are
thickness, permittivity and permeability of the layers,
respectively.

The calculations are performed using the characteristic matrix
method \cite{re36}, which is the most effective technique to analyze
the transmission properties of PCs. The characteristic matrix for
$(AB)^{N/2} C (AB)^{N/2}$, and symmetric $(AB)^{N/2} C (BA)^{N/2}$
structures, are given by:\\
$M[d]=(M_{A} M_{B})^{N/2} M_{C} (M_{A} M_{B})^{N/2}$, and $M[d]=(M_{A} M_{B})^{N/2} M_{C} (M_{B} M_{A})^{N/2}$, \\
where $M_{A}$, $M_{B}$ and $M_{C}$ are the characteristic matrices
of layers \emph{A}, \emph{B}, and \emph{C}. The characteristic
matrix $M_{i}$ for TE wave is given by \cite{re36}:
\begin{equation}
M_i= \left[
\begin{array}{cc}
\cos \gamma_{i}  &  \frac{-i}{p_{i}} \ \sin \gamma_{i}\\
-i \ p_{i} \ \sin \gamma_{i}  &  \cos \gamma_{i}
\end{array}\right]
\end{equation}
where, $\gamma_{i}=(\omega /c) \: n_{i} d_{i} \cos\theta_{i} $,
\emph{c} is speed of light in vacuum, $\theta_{i}$ is the ray angle
inside the layer \emph{i} with refractive index $n_i$,
$p_{i}=\sqrt{\varepsilon_{i}/ \mu_{i}}\:\cos\theta_{i}$, and
$\cos\theta_{i}=\sqrt{1-(n_{0}^2\:\sin^2\theta_{0}/{n_{i}^2})}$, in
which $n_0$ is the refractive index of the environment wherein the
incidence wave tends to enter the structure. The refractive index is
given as $n_i=\pm \sqrt{\varepsilon_{i}\mu_{i}}$ \cite{re31,re37},
where the positive and the negative signs are assigned for the PIM
and NIM layers, respectively.

The final characteristic matrix for an \emph{N} period structure is
given by:
\begin{equation}
[M(d)]^N= \prod^N_{i=1} M_i \equiv \left(
\begin{array}{cc}
m_{11}  &  m_{12}\\
m_{21}  &  m_{22}
\end{array}\right),
\end{equation}
where $m_{i,j}(i,j=1,2)$ are the matrix elements of $[M(d)]^N$. The transmission coefficient of the multilayer is calculated by:
\begin{equation}
t=\frac{2\ p_{0}}{(m_{11}+m_{12}\ p_{s})\ p_{0}+(m_{21}+m_{22}\
p_{s})}.
\end{equation}

In this equation, $p_{0}=n_{0}/ \cos\theta_{0}$ and $p_{s}=n_{s}/
\cos\theta_{s}$, with $n_s$ being the refractive index of the
environment where the wave leaves the crystal with angle
$\theta_{s}$. The transmissivity of the multilayer is given by
$T=(p_{s}/p_{0}) |t|^2 $. The transmissivity of the multilayer for
TM waves can be obtained by using these previous expressions with
$p_{i}=\sqrt{\mu_{i}/\varepsilon_{i}}\cos\theta_{i}$,
$p_{0}=\cos\theta_{0} / n_{0}$, and $p_{s}=\cos\theta_{s} / n_{s}$.

As mentioned before, our study is based on two different types of
NIMs. The permittivity and permeability of type-I NIM layer with
negative refracting index in the microwave region are complex, and
are defined as \cite{re38,re39},
\begin{equation}
\label{nineeq}
      \varepsilon(f) =1+\frac{5^2}{0.9^2-f^2-i\gamma f}+\frac{10^2}{11.5^2-f^2-i\gamma f}
\end{equation}
\begin{equation}
\label{teneq}
      \mu(f) =1+\frac{3^2}{0.902^2-f^2-i\gamma f}
\end{equation}
where \emph{f} and $\gamma$ are frequency and damping frequency in
GHz, respectively. As mentioned in \cite{re39}, for $f<3.13$ GHz,
the real parts of the permittivity and permeability,
$\varepsilon^{'}$ and $\mu^{'}$, are simultaneously negative
(double-negative material). For $3.13<f<3.78$ GHz,
$\varepsilon^{'}<0$, but $\mu^{'}>0$ (single-negative material
extending to the epsilon-negative material), and also for $f>3.78$
GHz, both $\varepsilon^{'}$ and $\mu^{'}$ are positive
(double-positive material). For type-II NIM layer, we use the Drude
model \cite{re40} to describe the complex permittivity and
permeability, with the results
\begin{equation}
      \varepsilon(f) =1-\frac{100}{f^2-i\gamma f}
\end{equation}
\begin{equation}
      \mu(f) =1.44-\frac{100}{f^2-i\gamma f}.
\end{equation}

Similarly, \emph{f} and $\gamma$ are respectively frequency and
damping frequency, given in GHz. For $f<8.33$ GHz, both
$\varepsilon^{'}$ and $\mu^{'}$, the real parts of the permittivity
and permeability, are negative (double-negative material), and for
$8.33<f<10$ GHz, $\varepsilon^{'}>0$, but $\mu^{'}<0$ (mu-negative
material).

\section{Numerical results and discussion}

Based on the theoretical model described in previous section, the
transmission spectrum of the lossy defective PC with a NIM defect
layer at the center is calculated. The calculations are carried out
in the region where the real parts of the permittivity and
permeability of two types of NIM layers (layers \emph{A}, and
\emph{C}), $\varepsilon^{'}$ and $\mu^{'}$, are simultaneously
negative where the zero-$\bar{n}$ gap will appear
\cite{re38,re39,re41,re42}. Equations (4) and (5), type-I NIM, are
used for the permittivity and permeability of layer A. PIM layer
(layer \emph{B}) is assumed to be the vacuum layer with $n_B=1$. The
thickness of layers \emph{A}, \emph{B}, and \emph{C} are
respectively chosen as $d_A=6$ mm, $d_B=12$ mm, and $d_C=24$ mm.
Also, the total number of the lattice period is selected to be
$N=16$ \cite{re39}.

In the first part, we use equations (4) and (5), type-I NIM, for the
permittivity and the permeability of NIM defect layer (layer
\emph{C}) exactly like layer \emph{A}, and the transmission spectra
of two different asymmetric and symmetric geometric structures for
both polarizations are investigated. First of all, the transmission
spectra of TE and TM polarized waves for the asymmetric structure at
various angles of incidence and for $\gamma = 0.2\times10^{-3}$ GHz
are shown in Figures 2 and 3. As it is seen, a single resonant peak
appears within the PBG, which corresponds to the type-I NIM defect
layer. This is in line with the report by Wu et al. \cite{re23} for
the defect modes in 1D PCs with PIMs where only one defect mode for
the asymmetric structure has been observed. In addition, the figures
show that as the angle of incidence increases, the peak height of
the defect mode decreases for TE waves and increases for TM waves.
Moreover, for both polarizations the frequency of the defect mode is
shifted to the higher frequency as the angle increases.

In Figures 4 and 5 we have plotted the frequency-dependent
transmittance for the symmetric structure in TE and TM waves at four
different angles of incidence. As it is observed in these figures,
similar to the asymmetric structure, there is a single defect mode
within the band gap again. This is in sharp contrast to the work by
Wu et al. \cite{re23} for the defect modes in 1D symmetric PC with
PIMs where there are two defect modes. It is also worth mentioning
that the number of defect modes appearing with type-I NIM defect
layer does not depend on the geometric structure, polarization, and
angle of the incidence wave.

In more detail, as shown in Figure 6, the frequency of defect mode
versus the angle of incidence for asymmetric (Figure 6(a)) and
symmetric (Figure 6(b)) structures have been respectively plotted.
As we mentioned earlier, only one defect mode appears in both
asymmetric and symmetric structures. Additionally, as seen from the
figures, in the asymmetric structure the defect modes appear in
higher frequencies compared to the symmetric one. Another feature in
Figure 6 which is worth mentioning is the frequency of the defect
mode, which, in TE waves, remains nearly unchanged in both geometric
structures as the angle of incidence increases. However, in TM
waves, the frequency increases as the angle of incidence increases.
In addition, this increase in symmetric structure is higher than
that of asymmetric one.

In the second part, properties of the defect mode of both geometric
structures with another type of NIM defect layer at the center
(type-II NIM), which was described in Section 2, are investigated.
The asymmetric and symmetric structures which were used before are
modified by replacing type-II NIM defect layer (layer \emph{C}), in
which the permittivity and permeability follow the equations (6) and
(7). The other parameters are kept the same as the first part.

In Figures 7 and 8, we have respectively plotted the transmission
spectra of TE and TM waves for the asymmetric structure at various
angles of incidence, when the loss factors of two types of NIMs,
layer \emph{A} and \emph{C}, are equal to $\gamma =
0.2\times10^{-3}$ GHz. We observe that there are three defect modes
for both TE and TM waves, as identified by numbers 1, 2, and 3,
respectively. Additionally, the transmission spectra of TE and TM
waves for the symmetric structure have been plotted in Figures 9 and
10, respectively. In the symmetric structure, similar to the
asymmetric one, there are three defect modes again. It can be seen
from Figures 7 to 10 that the frequencies of the defect modes are
shifted to the higher frequency as the incidence angle increases.
Besides, these figures show that for both asymmetric and symmetric
structures, three defect modes appear, which are independent of the
geometry of the structure. To the contrary, as mentioned by Wu et
al. \cite{re23}, in the conventional PCs case for a PIM defect
layer, the symmetric geometric structure shows two defect modes and
the asymmetric one shows one defect mode.

In more detail, we have plotted the defect mode frequency versus
angle of incidence for both asymmetric (Figure 11(a)) and symmetric
(Figure 11(b)) structures. As seen from these figures, in TE waves
for both geometric structures, the frequency of defect modes remains
nearly unchanged with increasing the angle of incidence. On the
other hand, frequency of defect modes in TM waves increases as the
angle of incidence increases. Additionally, similar to the first
part, in the asymmetric structure the defect modes appear in higher
frequencies compared to the symmetric one. In contrast to type-I NIM
defect layer, frequencies of defect modes, which increase in the
asymmetric structure, are higher than that of the symmetric
structure. Another feature in Figure 11 is that in the asymmetric
structure, defect modes 1 and 2 disappear for angles more than
$45^\circ$ and $65^\circ$, respectively. Also, defect modes 1 and 2
disappear for angles more than $55^\circ$ and $70^\circ$,
respectively, in the symmetric structure. To the contrary, Figure 6,
by comparing defect mode 3 in both geometric structures for TM
waves, shows that the frequency of the defect mode in the asymmetric
structure changes more than that of the symmetric one.

As we observe, in comparison to the first part, which is type-I NIM
defect layer, the number of defect modes depends on the types of
NIMs which are being used for the defect layer. In addition, as the
figures show, for two types of NIMs, the number of defect modes does
not depend on the geometric structure. This is in sharp contrast
with the work done by Wu et al. \cite{re23} where the number of
defect modes depends on the asymmetric or symmetric structure.

\section{Conclusion}

This paper has theoretically investigated the properties of defect
modes coming from two types of NIMs defect layer at the center of 1D
defective lossy MPCs in symmetric and asymmetric geometric
structures. Our numerical results show that there exists a single
defect mode inside the PBG in both asymmetric and symmetric
structures for TE and TM polarized waves, when the type-I NIM defect
layer is considered. On the other hand, the results show that with
changing the types of NIMs in the defect layer, we observe three
defect modes inside the band gap for both structures. Positions
(frequencies) of the defect modes depend on the polarization of the
waves and the incidence angles. As the incidence angle increases,
the frequency of the defect modes moves toward higher frequencies
regardless of the geometrical structures and types of NIMs defect
layer. Additionally, the results of this study bring up the
conclusion that the number of defect modes in 1D asymmetric and
symmetric lossy PCs strongly depends on the types of NIM defect.
Besides, an important result is that by using an NIM in the
structure of PCs, the defect modes show different behavior from that
of the PCs with only PIMs. Finally, it is worthwhile to mention that
the number of defect modes does not depend on the asymmetric or
symmetric structures, contrary to the work conducted by Wu et al.
\cite{re23}. Detailed analysis of the defect modes in metamaterial
photonic crystals with different geometric structures will certainly
provide useful information for designing new types of narrowband and
multichannel transmission filters.

\acknowledgements

A. Aghajamali would like to acknowledge his gratitude to P. Shams
for her help and useful discussion.

%----------------------------------------------------------
\newpage
\thispagestyle{empty}
\begin{figure}[tbp]
\epsfxsize=7cm \centerline{\includegraphics [width=9cm] {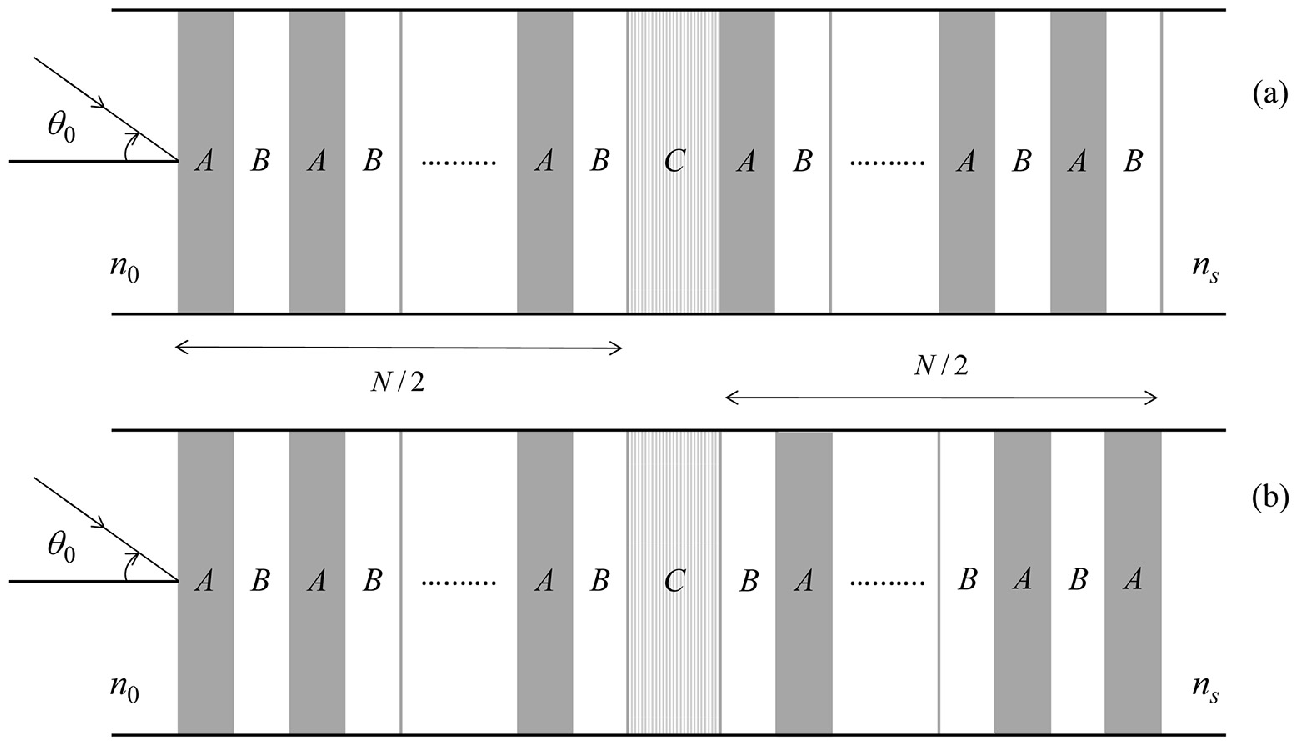}}
\caption{Schematic diagrams of 1D defective MPCs embedded in air,
with a defect layer, (a) asymmetric structure, (b) symmetric
structure, where layers \emph{A} and \emph{C} are NIMs and layer
\emph{B} is PIM. The number of the periods is \emph{N} and the
incidence angle is $\theta_{0}$.}
\end{figure}

\begin{figure}[tbp]
\epsfxsize=7cm \centerline{\includegraphics [width=11cm] {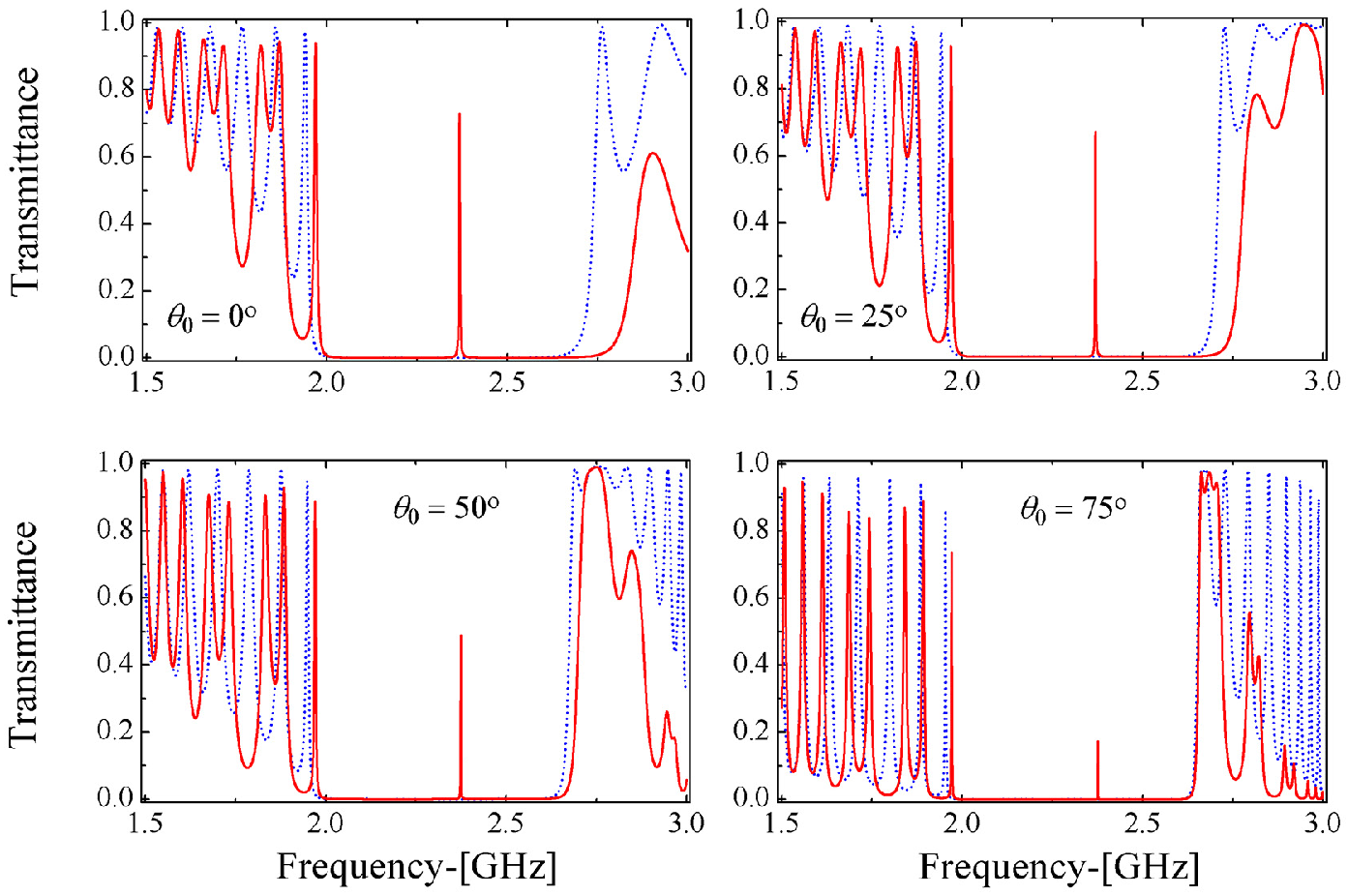}}
\caption{Transmission spectra of TE polarized wave for the
asymmetric 1D MPC structures with type-I NIM defect layer (solid
line) and without defect layer (dotted line) for different incidence
angles with $\gamma = 0.2\times10^{-3}$ GHz.}
\end{figure}

\begin{figure}[tbp]
\epsfxsize=7cm \centerline{\includegraphics [width=11cm] {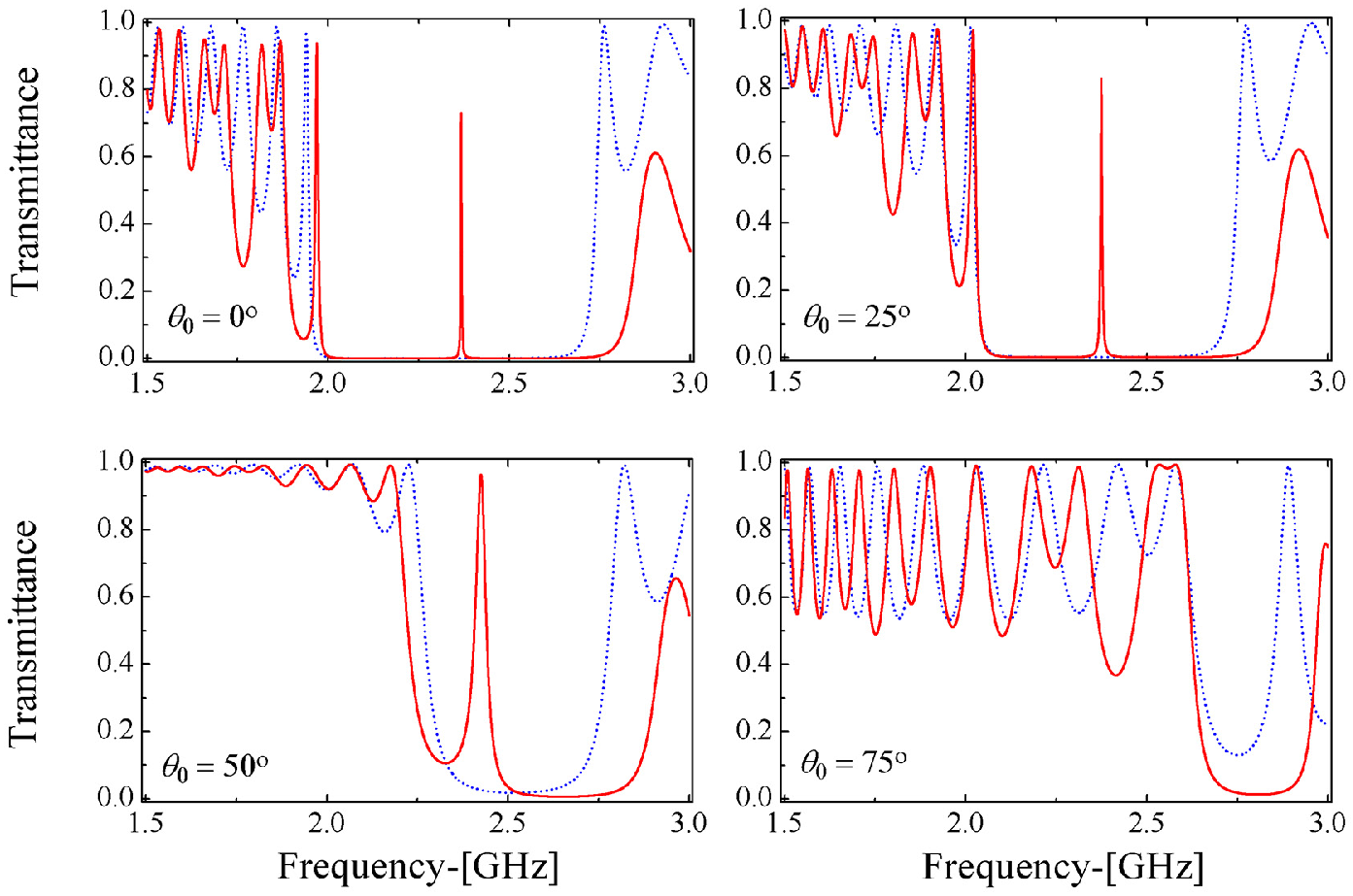}}
\caption{Transmission spectra of TM polarized wave for the
asymmetric 1D MPC structures with type-I NIM defect layer (solid
line) and without defect layer (dotted line) for different incidence
angles with $\gamma = 0.2\times10^{-3}$ GHz.}
\end{figure}

\begin{figure}[tbp]
\epsfxsize=7cm \centerline{\includegraphics [width=11cm] {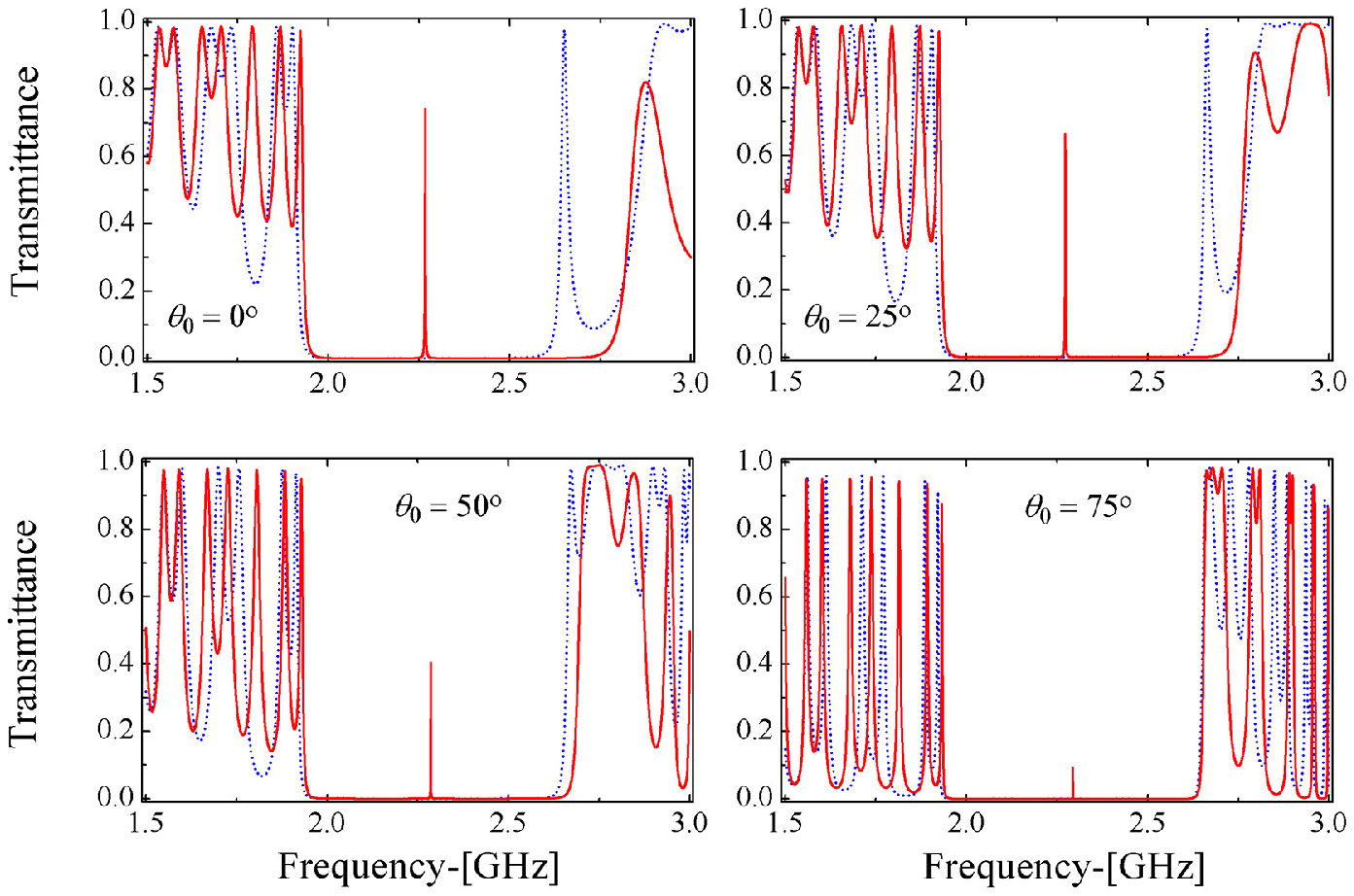}}
\caption{Transmission spectra of TE polarized wave for the symmetric
1D MPC structures with type-I NIM defect layer (solid line) and
without defect layer (dotted line) for different incidence angles
with $\gamma = 0.2\times10^{-3}$ GHz.}
\end{figure}

\begin{figure}[tbp]
\epsfxsize=7cm \centerline{\includegraphics [width=11cm] {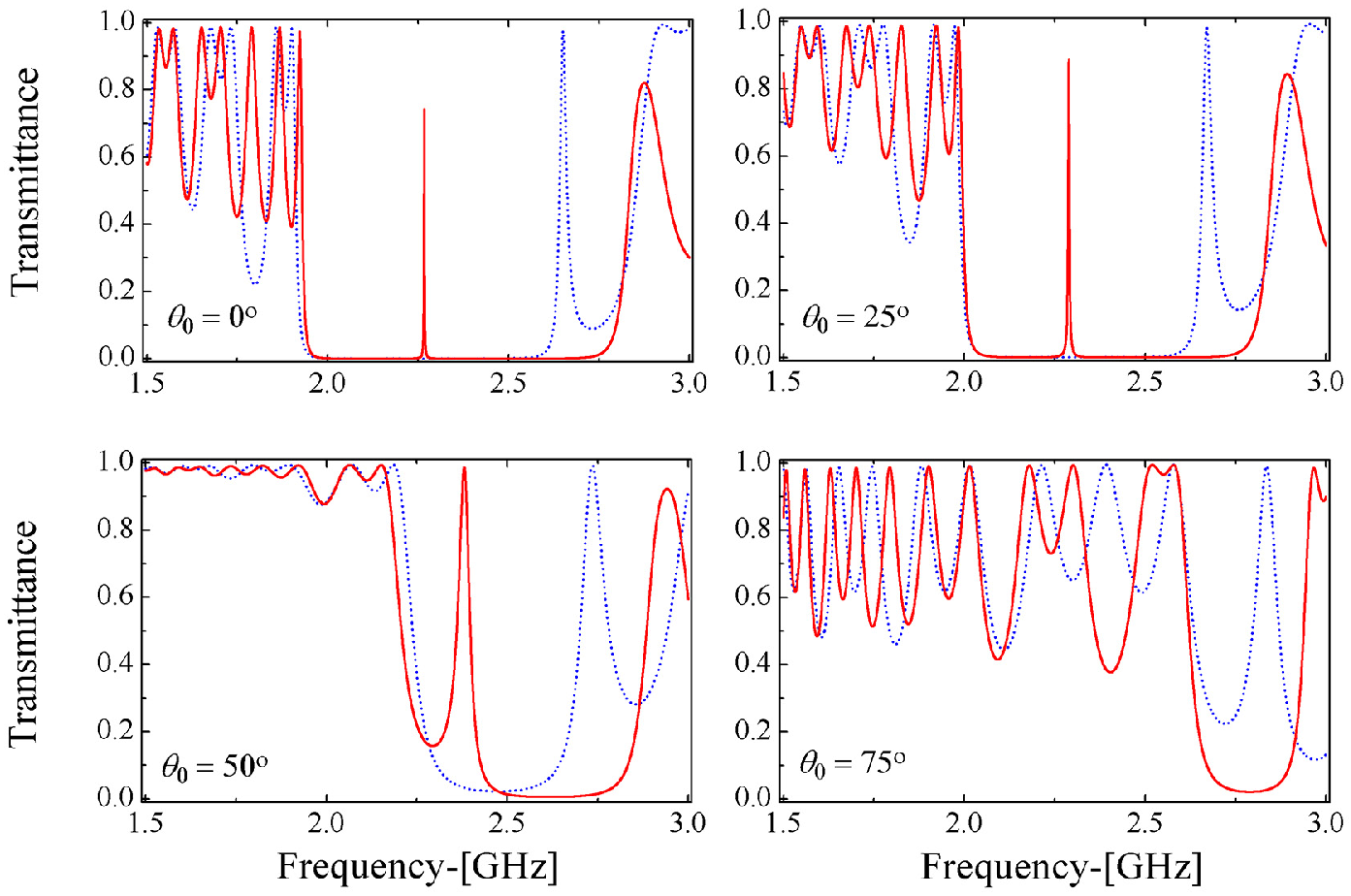}}
\caption{Transmission spectra of TM polarized wave for the symmetric
1D MPC structures with type-I NIM defect layer (solid line) and
without defect layer (dotted line) for different incidence angles
with $\gamma = 0.2\times10^{-3}$ GHz.}
\end{figure}

\begin{figure}[tbp]
\epsfxsize=7cm \centerline{\includegraphics [width=11cm] {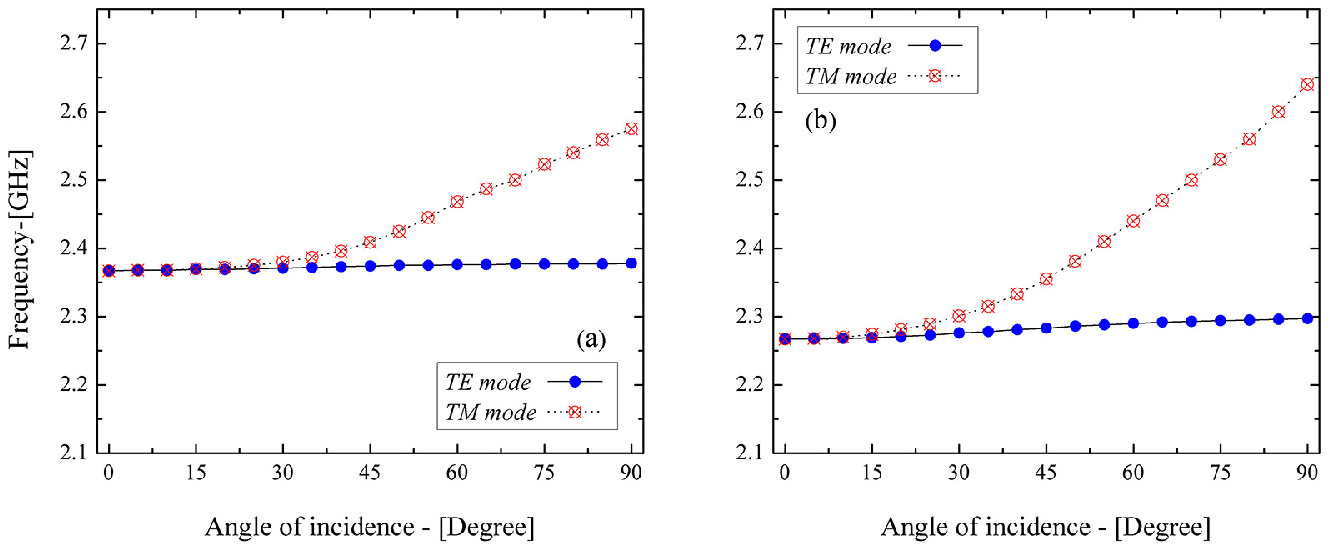}}
\caption{Frequency of the defect modes in (a) the asymmetric and (b)
the symmetric 1D PMCs as a function of angle of incidence for both
polarizations, with $\gamma = 0.2\times10^{-3}$ GHz.}
\end{figure}

\begin{figure}[tbp]
\epsfxsize=7cm \centerline{\includegraphics [width=11cm] {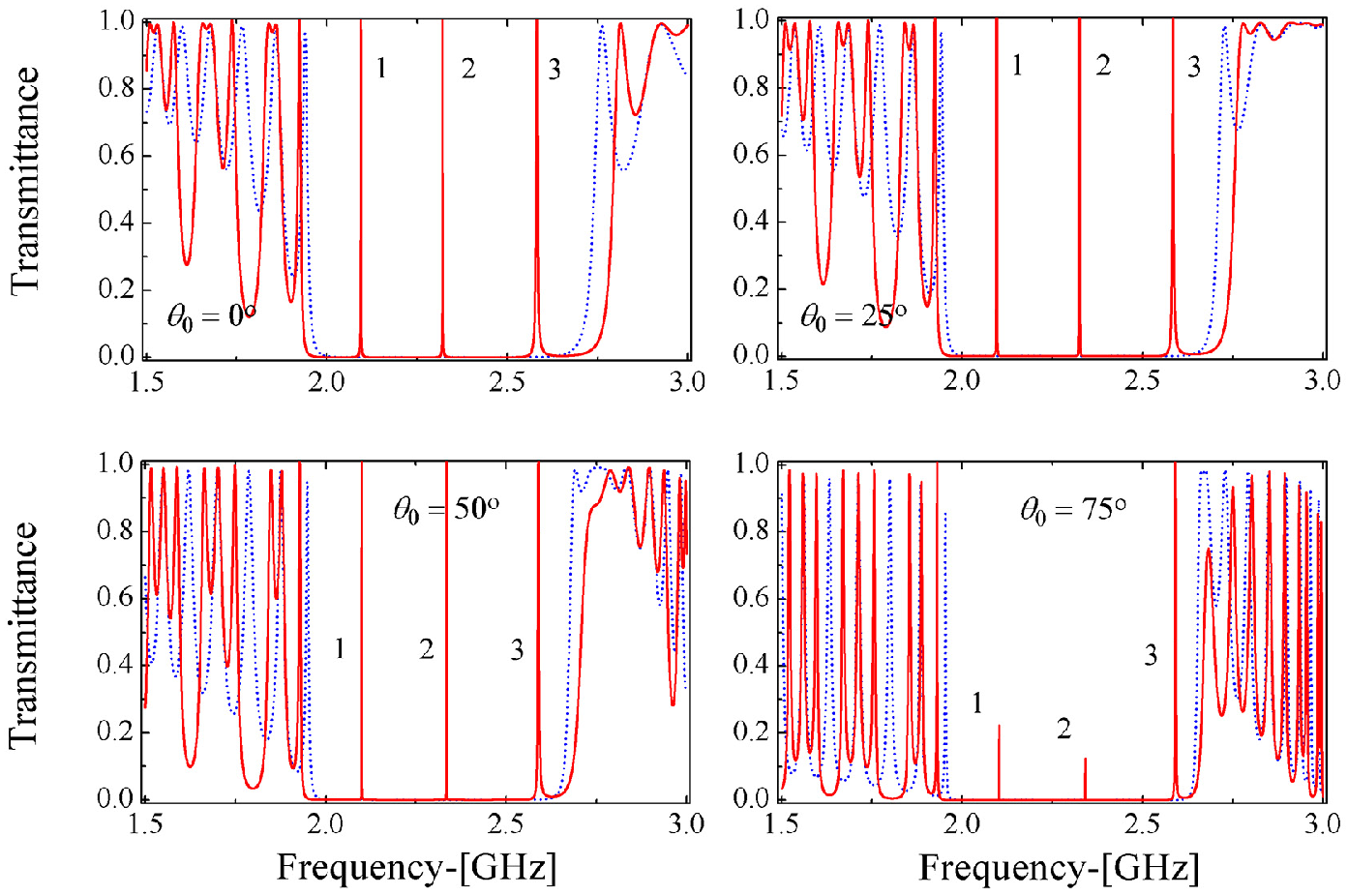}}
\caption{Transmission spectra of TE polarized wave for the
asymmetric 1D MPC structures with type-II NIM defect layer (solid
line) and without defect layer (dotted line) for different incidence
angles with $\gamma = 0.2\times10^{-3}$ GHz.}
\end{figure}

\begin{figure}[tbp]
\epsfxsize=7cm \centerline{\includegraphics [width=11cm] {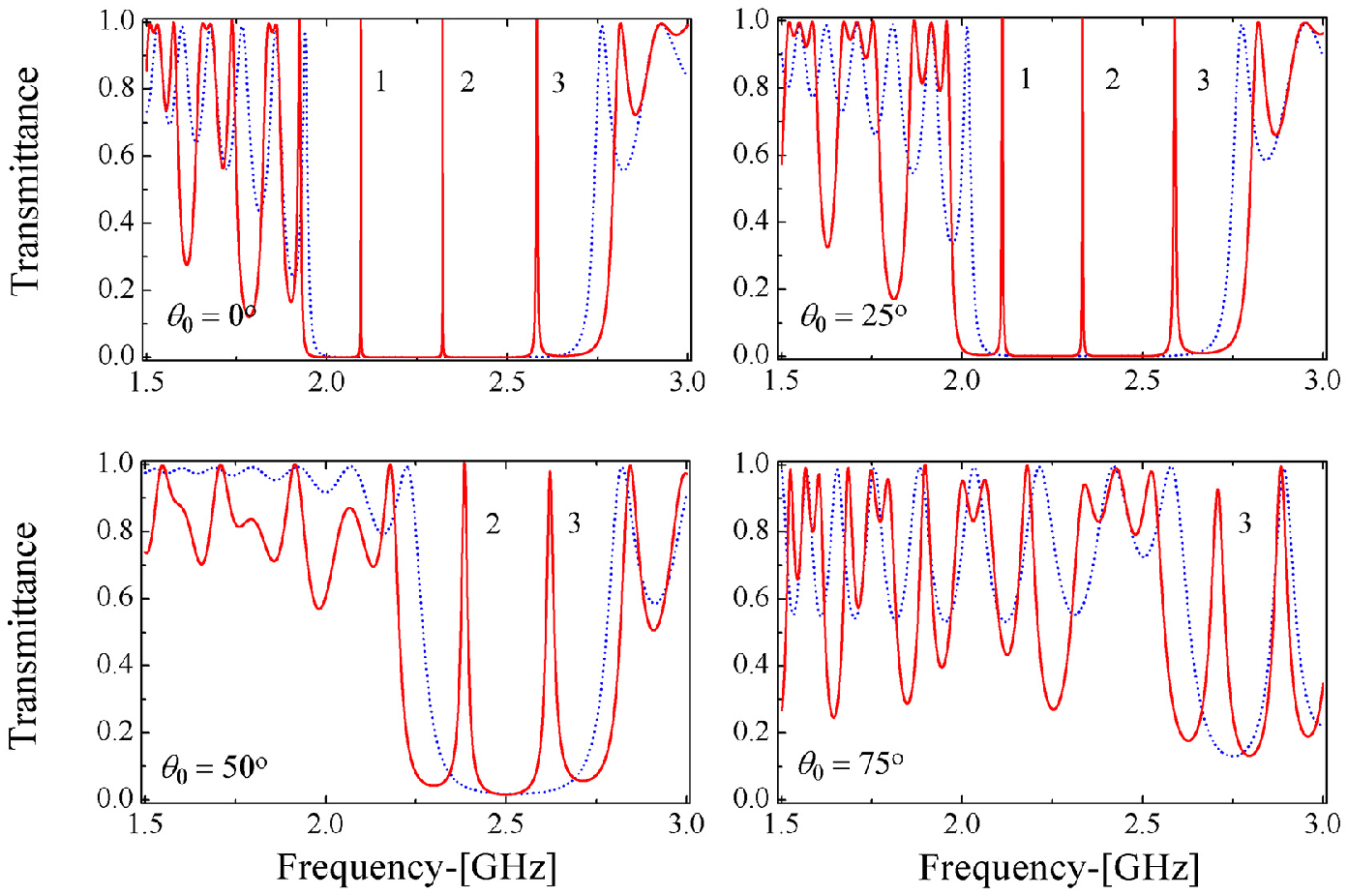}}
\caption{Transmission spectra of TM polarized wave for the
asymmetric 1D MPC structures with type-II NIM defect layer (solid
line) and without defect layer (dotted line) for different incidence
angles with $\gamma = 0.2\times10^{-3}$ GHz.}
\end{figure}

\begin{figure}[tbp]
\epsfxsize=7cm \centerline{\includegraphics [width=11cm] {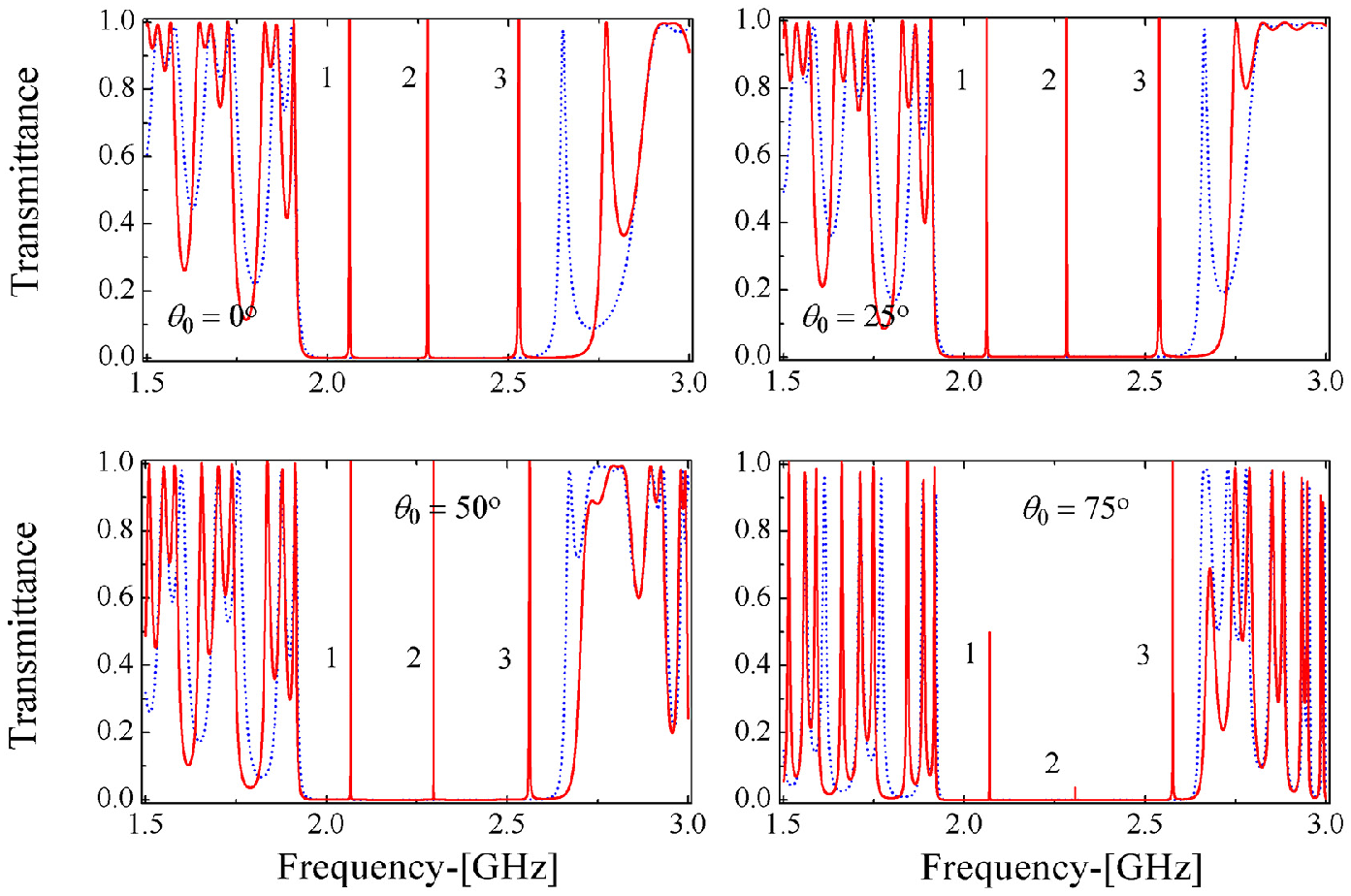}}
\caption{Transmission spectra of TE polarized wave for the symmetric
1D MPC structures with type-II NIM defect layer (solid line) and
without defect layer (dotted line) for different incidence angles
with $\gamma = 0.2\times10^{-3}$ GHz.}
\end{figure}

\begin{figure}[tbp]
\epsfxsize=7cm \centerline{\includegraphics [width=11cm] {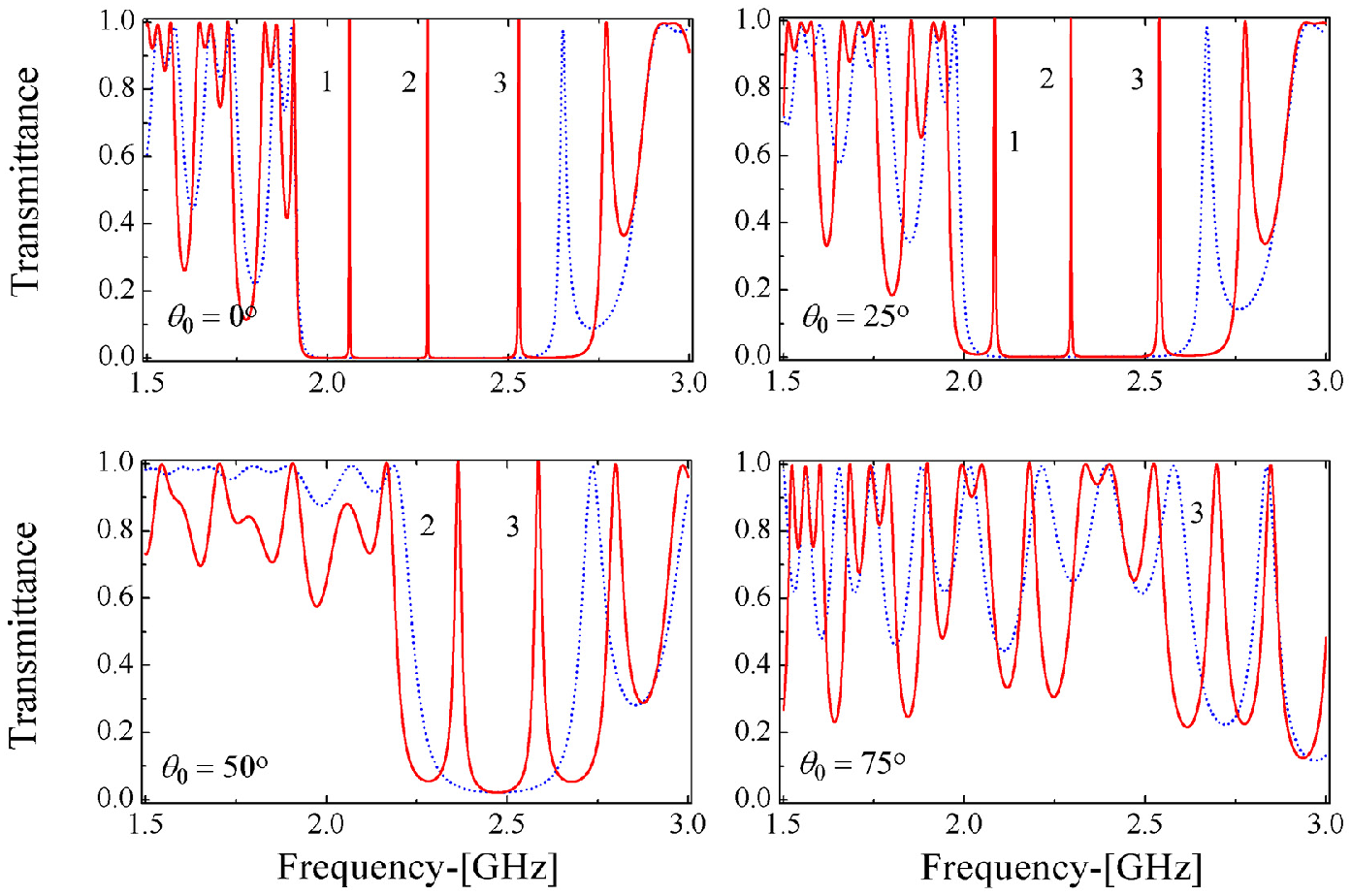}}
\caption{Transmission spectra of TM polarized wave for the symmetric
1D MPC structures with type-II NIM defect layer (solid line) and
without defect layer (dotted line) for different incidence angles
with $\gamma = 0.2\times10^{-3}$ GHz.}
\end{figure}

\begin{figure}[tbp]
\epsfxsize=7cm \centerline{\includegraphics [width=11cm] {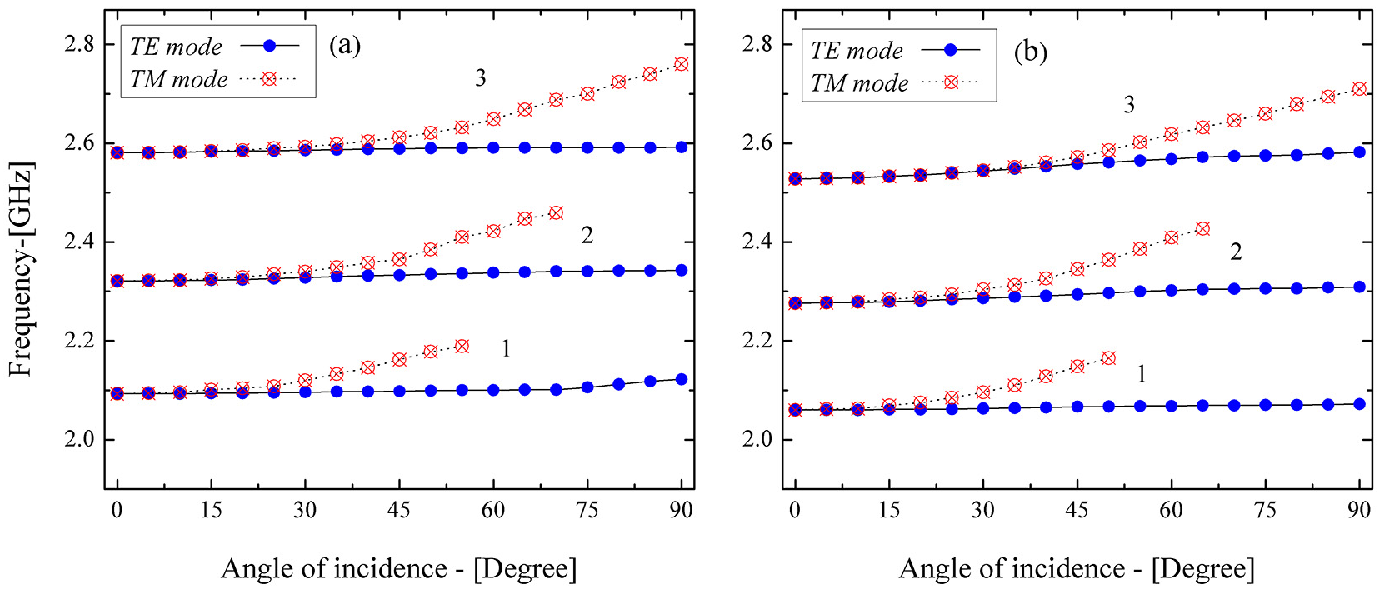}}
\caption{Frequency of the defect modes in (a) the asymmetric and (b)
the symmetric 1D PMCs as a function of angle of incidence for both
polarizations, with $\gamma = 0.2\times10^{-3}$ GHz.}
\end{figure}
%----------------------------------------------------------

\end{document}